\documentclass[range]{ar2e-changed}
\begin{document}
\renewcommand\d{\partial}
\newcommand\x{\mathbf{x}}
\renewcommand\k{\mathbf{k}}
\newcommand\<{\langle}
\renewcommand\>{\rangle}
\newcommand\Tr{\mathop{\mathrm{Tr}}}
\renewcommand\Im{\mathop{\mathrm{Im}}}
\newcommand{\ti}{{t_{\mathrm{i}}}}
\newcommand{\tf}{{t_{\mathrm{f}}}}

\input epsf.tex    

\input psfig.sty

\ARinfo{\hspace{5in}INT PUB 07-02}

\title{Viscosity, Black Holes, and Quantum Field Theory}

\markboth{Son, Starinets}{Viscosity, Black Holes, and QFT}

\author{Dam T. Son
\affiliation{Institute for Nuclear Theory, University of Washington,
Seattle, Washington 98195-1550, USA}
Andrei O. Starinets
\affiliation{Perimeter Institute for Theoretical Physics, Waterloo, Ontario
N2L 2Y5, Canada}}


\begin{abstract}
We review recent progress in applying the AdS/CFT correspondence to
finite-temperature field theory.  In particular, we show how the
hydrodynamic behavior of field theory is reflected in the low-momentum
limit of correlation functions computed through a real-time AdS/CFT
prescription, which we formulate.  We also show how the hydrodynamic
modes in field theory correspond to the low-lying quasinormal modes of
the AdS black p-brane metric.  We provide a proof of the universality
of the viscosity/entropy ratio within a class of theories with gravity
duals and formulate a viscosity bound conjecture.  Possible
implications for real systems are mentioned.
\end{abstract}

\maketitle

\section{INTRODUCTION}

This review is about the recently emerging connection, through the
gauge/gravity correspondence, between hydrodynamics and black hole
physics.

The study of quantum field theory at high temperature has a long
history.  It was first motivated by the Big Bang cosmology when it was
hoped that early phase transitions might leave some imprints on the
Universe~\cite{Kirzhnits:1972ut}.  One of those phase transitions is
the QCD phase transitions (which could actually be a crossover) which
happened at a temperature around $T_c\sim 200$~MeV, when matter turned
from a gas of quarks and gluons (the quark-gluon plasma, or QGP) into
a gas of hadrons.

An experimental program was designed to create and study the QGP by
colliding two heavy atomic nuclei.  Most recent experiments are
conducted at the Relativistic Heavy Ion Collider (RHIC) at Brookhaven
National Laboratory.  Although significant 
 circumstantial evidence for
the QGP was accumulated~\cite{Gyulassy:2004zy}, a theoretical
interpretation of most of the experimental data proved difficult, because
the QGP created at RHIC is far from
being a weakly coupled gas of quarks and gluons.  Indeed, the
temperature of the plasma, as inferred from the spectrum of final
particles, is only approximately 170 MeV, near the confinement scale of QCD.
This is deep in the nonperturbative regime of QCD, where reliable
theoretical tools are lacking.  Most notably, the kinetic coefficients
of the QGP, which enter the hydrodynamic equations (reviewed in
Sec.~\ref{sec:hydro}), are not theoretically computable at these
temperatures.

The paucity of information about the kinetic coefficients of the QGP in
particular and of strongly coupled thermal quantum field theories in
general is one of the main reasons for our interest in their
computation in a class of strongly coupled field theories, even though
this class does not include QCD.  The necessary technological tool is
the anti--de Sitter--conformal field theory (AdS/CFT)
correspondence~\cite{Maldacena:1997re,Gubser:1998bc,Witten:1998qj},
discovered in the investigation of D-branes in string theory.  This
correspondence allows one to describe the thermal plasmas in these
theories in terms of black holes in AdS space.  The AdS/CFT
correspondence is reviewed in Sec.~\ref{sec:AdSCFT}.

The first calculation of this type, that of the shear viscosity in
${\cal N}=4$ supersymmetric Yang-Mills (SYM)
theory~\cite{Policastro:2001yc}, is followed by the theoretical work
to establish the rules of real-time finite-temperature AdS/CFT
correspondence~\cite{Son:2002sd,Herzog:2002pc}.  Applications of
these rules to various special
cases~
\cite{Policastro:2002se,Policastro:2002tn,Herzog:2002fn,Herzog:2003ke}
clearly show that even very exotic field theories, when heated up to
finite temperature, behave hydrodynamically at large distances and
time scales (provided that the number of space-time dimensions is
$2{+}1$ or higher).  This development is reviewed in
Sec.~\ref{sec:realtime}.  Moreover, the way AdS/CFT works reveals       
deep connections to properties of black holes in classical gravity.
For example, the hydrodynamic modes of a thermal medium are mapped,
through the correspondence, to the low-lying quasi-normal modes of a 
black-brane metric.  It seems that our understanding of the connection
between hydrodynamics and black hole physics is still incomplete;
we may understand more about gravity
by studying thermal field theories.  One idea along this direction is
reviewed in Sec.~\ref{sec:membrane}.

From the point of view of heavy-ion (QGP) physics, a
particularly interesting finding has been the formulation of a
conjecture on the lowest possible value of the ratio of viscosity and
volume density of entropy.  This conjecture was motivated by the
universality of this ratio in theories with gravity duals.  This is
reviewed in Sec.~\ref{sec:ratio}.

This review is written primarily for readers with a background in QCD
and QGP physics who are interested in learning about
AdS/CFT correspondence and its applications to finite-temperature
field theory.  Some parts of this review (for example, the section
about hydrodynamics) should be useful for readers with a string theory
or general relativity background who are interested in the connection
between string theory, gravity, and hydrodynamics.  The perspectives
here are shaped by our personal taste and therefore may appear
narrow, but the authors believe that this review may serve as the
starting point to explore the much richer original literature.

In this review we use the ``mostly plus'' metric signature $-+++$.

\section{HYDRODYNAMICS}
\label{sec:hydro}

From the modern perspective, hydrodynamics~\cite{Forster} is best
thought of as an effective theory, describing the dynamics at large
distances and time-scales.  Unlike the familiar effective field
theories (for example, the chiral perturbation theory), it is normally
formulated in the language of equations of motion instead of an action
principle.  The reason for this is the presence of dissipation in
thermal media.

In the simplest case, the hydrodynamic equations are just the laws of
conservation of energy and momentum,
\begin{equation}
  \d_\mu T^{\mu\nu} = 0\,.
\end{equation} 
To close the system of equations, we must reduce the number of independent
elements of $T^{\mu\nu}$.  This is done through the assumption of
\emph{local thermal equilibrium:} If perturbations have long wavelengths,
the state of the system, at a given time, is determined by the
temperature as a function of coordinates $T(\x)$ and the local fluid
velocity $u^\mu$, which is also a function of coordinates
$u^\mu(\x)$.  Because $u_\mu u^\mu=-1$, only three components of $u^\mu$
are independent. The number of hydrodynamic variables is four, equal
to the number of equations.

In hydrodynamics we express $T^{\mu\nu}$ through $T(x)$ and $u^\mu(x)$
through the so-called constitutive equations.  Following the standard
procedure of effective field theories, we expand in powers of spatial
derivatives.  To zeroth order, $T^{\mu\nu}$ is given by the familiar
formula for ideal fluids,
\begin{equation}\label{Tmunu-ideal}
  T^{\mu\nu} = (\epsilon + P)u^\mu u^\nu + Pg^{\mu\nu}\,,
\end{equation}
where $\epsilon$ is the energy density, and $P$ is the pressure.  Normally
one would stop at this leading order, but 
qualitatively new effects necessitate going to the next order.
Indeed, from Eq.~\ref{Tmunu-ideal} and the thermodynamic relations
$d\epsilon=TdS$, $dP=sdT$, and $\epsilon+P=Ts$ ($s$ is the entropy per
unit volume), one finds that entropy is conserved~\cite{LL6}
\begin{equation}
  \d_\mu (su^\mu) = 0\,.
\end{equation}
Thus, to have entropy production, one needs to go to the next order in the
derivative expansion.

At the next order, we write
\begin{equation}
  T^{\mu\nu} =  (\epsilon + P)u^\mu u^\nu + Pg^{\mu\nu} -\sigma^{\mu\nu}\,,
\end{equation}
where $\sigma^{\mu\nu}$ is proportional to derivatives of $T(x)$ and
$u^\mu(x)$ and is termed the dissipative part of $T^{\mu\nu}$.  To
write these terms, let us first fix a point $x$ and go to the local rest
frame where $u^i(x)=0$.  In this frame, in principle one can have
dissipative corrections to the energy-momentum density $T^{0\mu}$.
However, one recalls that the choice of $T$ and $u^\mu$ is arbitrary,
and thus one can always redefine them so that these corrections vanish,
$\sigma^{00}=\sigma^{0i}=0$, and so at a point $x$,
\begin{equation}
  T^{00} = \epsilon, \qquad T^{0i} = 0\,.
\end{equation}
The only nonzero elements of the dissipative energy-momentum tensor
are $\sigma_{ij}$.  To the next-to-leading order there are extra
contributions whose forms are dictated by rotational symmetry:
\begin{equation}
  \sigma_{ij} =  \eta\left(\d_i u_j + \d_j u_i 
     - \frac 23\delta_{ij}\d_k u^k\right) + \zeta\delta_{ij} \d_k u^k\,.
\end{equation}
Going back to the general frame, we can now write the dissipative part
of the energy-momentum tensor as
\begin{equation}\label{sigma-PP}
  \sigma^{\mu\nu} = P^{\mu\alpha} P^{\nu\beta} 
  \left[\eta\left(\d_\alpha u_\beta + \d_\beta u_\alpha 
  - \frac23g_{\alpha\beta}
    \d_\lambda u^\lambda \right) 
    + \zeta g_{\alpha\beta} \d_\lambda u^\lambda\right]\,,
\end{equation}
where $P^{\mu\nu}=g^{\mu\nu}+u^\mu u^\nu$ is the projection operator
onto the directions perpendicular to $u^\mu$.

If the system contains a conserved current, there is an
additional hydrodynamic equation related to the current conservation,
\begin{equation}
  \d_\mu j^\mu = 0\,.
\end{equation}
The constitutive equation contains two terms:
\begin{equation}
  j^\mu = \rho u^\mu - DP^{\mu\nu}\d_\nu\alpha\,,
\end{equation}
where $\rho$ is the charge density in the fluid rest frame and 
$D$ is some constant.  The first
term corresponds to convection, the second one to diffusion.  In the
fluid rest frame, ${\bf j}=-D\nabla\rho$, which is Fick's law of
diffusion, with $D$ being the diffusion constant.

\subsection{Kubo's Formula For Viscosity}

As mentioned above, the hydrodynamic equations can be thought
of as an effective theory describing the dynamics of the system at
large lengths and time scales.  Therefore one should be able to use these
equations to extract information about the low-momentum behavior of
Green's functions in the original theory.

For example, let us recall how the two-point correlation
functions can be extracted.  If we couple sources $J_a(\x)$ to a set
of (bosonic) operators $O_a(x)$, so that the new action is
\begin{equation}
  S = S_0 + \int_x J_a(x) O_a(x)\,,
\end{equation}
then the source will introduce a perturbation of the system.  In
particular, the average values of $O_a$ will differ from the
equilibrium values, which we assume to be zero.  If $J_a$ are small, the
perturbations are given by the linear response theory as
\begin{equation}
  \<O_a(x)\> = -\int_y G^R_{ab}(x-y) J_b(y)\,,
\end{equation}
where $G^R_{ab}$ is the retarded Green's function
\begin{equation}
  iG^R_{ab}(x-y) = \theta(x^0-y^0) \<[O_a(x),\, O_b(y)]\> \,.
\end{equation}
The fact that  the linear response is determined by 
the retarded (and not by any other)  
Green's function is obvious from
causality: The source can influence the system only after it has been 
 turned on.

Thus, to determine the correlation functions of $T^{\mu\nu}$,
we need to couple a weak source to $T^{\mu\nu}$ and determine the
average value of $T^{\mu\nu}$ after this source is turned on.  To find
these correlators at low momenta, we can use the hydrodynamic theory.
So far in our treatment of hydrodynamics we have included no source
coupled to $T^{\mu\nu}$.  This deficiency can be easily
corrected, as the source of the energy-momentum tensor is the metric
$g_{\mu\nu}$.  One must generalize the hydrodynamic equations to
curved space-time and from it determine the response of the thermal
medium to a weak perturbation of the metric.  This procedure is rather
straightforward and in the interest of space is left as an exercise to
the reader.

Here we concentrate on a particular case when
the metric perturbation is homogeneous in space but time dependent:
\begin{eqnarray}\label{g-pert}
  g_{ij}(t,\x) &=& \delta_{ij} + h_{ij}(t), \qquad h_{ij}\ll 1\\
  g_{00}(t,\x) &=& -1, \qquad g_{0i}(t,\x) =0\,.
\end{eqnarray}
Moreover, we assume the perturbation to be traceless, $h_{ii}=0$.
Because the perturbation is spatially homogeneous, 
if the fluid moves, it
can only move uniformly: $u^i=u^i(t)$.  However, this possibility can be
ruled out by parity, so the fluid must remain at rest all the time:
$u^\mu=(1,0,0,0)$.  We now compute the dissipative part of the
stress-energy tensor.  The generalization of Eq.~\ref{sigma-PP} to
curved space-time is
\begin{equation}
  \sigma^{\mu\nu} = P^{\mu\alpha} P^{\nu\beta} 
  \left[ \eta (\nabla_\alpha u_\beta + \nabla_\beta u_\alpha)
  + \left(\zeta - \frac23\eta\right)g_{\alpha\beta}\nabla\cdot u\right]\,.
\end{equation}
Substituting $u^\mu=(1,0,0,0)$ and $g_{\mu\nu}$ from
Eq.~\ref{g-pert}, we find only contributions to the traceless
spatial components, and these contributions come entirely from the
Christoffel symbols in the covariant derivatives.  For example,
\begin{equation}
  \sigma_{xy} = 2\eta\Gamma^0_{xy} = \eta\d_0 h_{xy} \,.
\end{equation}
By comparison with the expectation from the linear response theory, this
equation means that we have found the zero spatial momentum,
low-frequency limit of the retarded Green's function of $T^{xy}$:
\begin{equation}
  G^R_{xy,xy}(\omega,{\bf 0}) = \int\!dt\,d\x\, e^{i\omega t} \theta(t) 
  \<[ T_{xy}(t,\x),\, T_{xy}(0,{\bf 0})]\> = -i\eta\omega + O(\omega^2)
\end{equation}
(modulo contact terms).  We have, in essence, derived the Kubo's
formula relating the shear viscosity and a Green's function:
\begin{equation}
  \eta = -\lim_{\omega\to0} \frac1\omega\Im G^R_{xy,xy}(\omega, {\bf 0})\,.
\label{kubo}
\end{equation}
There is a similar Kubo's relation for the charge diffusion constant
$D$.

\subsection{Hydrodynamic Modes}

If one is interested only in the locations of the poles of the
correlators, one can simply look for the normal modes of the
linearized hydrodynamic equations, that is, solutions that behave as
$e^{-i\omega t+i\k\cdot\x}$.  Owing to dissipation, the frequency
$\omega(\k)$ is complex.  For example, the equation of charge
diffusion,
\begin{equation}
  \d_t\rho - D\nabla^2 \rho =0,
\end{equation}
corresponds to a pole in the current-current correlator at
$\omega=-iDk^2$. 

To find the poles in the correlators between elements of the
stress-energy tensor one can, without loss of generality, choose the
coordinate system so that $\k$ is aligned along the $x^3$-axis:
$\k=(0,0,k)$.  Then one can distinguish two types of normal modes:

\begin{itemize}
\item[1.]\underline{Shear modes} correspond to the fluctuations of
pairs of components $T^{0a}$ and $T^{3a}$, 
where $a=1,2$.  The constitutive equation is
\begin{equation}
  T^{3a} = -\eta\d_3 u^a = -\frac\eta{\epsilon+P}\d_3 T^{0a}\,,
\end{equation}
and the equation for $T^{0a}$ is
\begin{equation}
  \d_t T^{0a} - \frac\eta{\epsilon+P} \d_3^2 T^{0a} = 0\,.
\end{equation}
That is, it has the form of a diffusion equation for $T^{0a}$.  Substituting
$e^{-i\omega t+ikx^3}$ into the equation, one finds the dispersion law
\begin{equation}
  \omega = -i \frac\eta{\epsilon+P} k^2\,.
\end{equation}

\item[2.]\underline{Sound modes} are fluctuations of 
$T^{00}$, $T^{03}$, and
$T^{33}$.  There are now two conservation equations, and by
diagonalizing them one finds the dispersion law
\begin{equation}
  \omega = c_s k -\frac i2
  \left(\frac43\eta+\zeta\right) \frac{k^2}{\epsilon+P}\,,
\end{equation}
where $c_s=\sqrt{dP/d\epsilon}$.  This is simply the sound wave, which
involves the fluctuation of the energy density.  It propagates with
velocity $c_s$, and its damping is related to a linear combination of
shear and bulk viscosities.
\end{itemize}

In CFTs it is possible to use conformal Ward
identities to show that the bulk viscosity vanishes: $\zeta=0$.  Hence,
we shall concentrate our attention on the shear viscosity $\eta$.

\subsection{Viscosity In Weakly Coupled Field Theories}
\label{visc:weak}

We now briefly consider the behavior of the shear viscosity in
weakly coupled field theories, with the $\lambda \phi^4$
 theory as a concrete
example.  At weak coupling, there is a separation between two length
scales: The mean free path of particles is much larger than the
distance scales over which scatterings occur.  Each scattering event
takes a time of order $T^{-1}$ (which can be thought of as the time
required for final particles to become on-shell).  The mean free path
$\ell_{\rm mfp}$ can be estimated from the formula
\begin{equation}
  \ell_{\rm mfp} \sim \frac1{n\sigma v}\,,
\end{equation}
where $n$ is the density of particles, $\sigma$ is the typical
scattering cross section, and $v$ is the typical particle velocity.
Inserting the values for thermal $\lambda \phi^4$ theory, $n\sim T^3$,
$\sigma\sim\lambda^2 T^{-2}$, and $v\sim 1$, one finds
\begin{equation}
  \ell_{\rm mfp} \sim \frac1{\lambda^2 T}\gg \frac1T\,.
\end{equation}

The viscosity can be estimated from kinetic theory to be
\begin{equation}
  \eta \sim \epsilon \ell_{\rm mfp}\,,
\end{equation}
where $\epsilon$ is the energy density.  From $\epsilon\sim T^4$ and
the estimate of $\ell_{\rm mft}$, one finds
\begin{equation}\label{eta-phi4-est}
  \eta \sim \frac{T^3}{\lambda^2}\,.
\end{equation}
In particular, the weaker the coupling $\lambda$, the larger the
viscosity $\eta$.  This behavior is explained by the fact that the
viscosity measures the rate of momentum diffusion.  The smaller
$\lambda$ is, the longer a particle travels before colliding with another
one, and the easier the momentum transfer.

It may appear counterintuitive that viscosity tends to infinity in the
limit of zero coupling $\lambda\to0$: At zero coupling there is no
dissipation, so should the viscosity be zero?  The confusion arises
owing to the fact that the hydrodynamic theory, and hence the notion of
viscosity, makes sense only on distances much larger than the mean
free path of particles.  If one takes $\lambda\to0$, then to measure
the viscosity one has to do the experiment at larger and larger length
scales.  If one fixes the size of the experiment and takes
$\lambda\to0$, dissipation disappears, but it does not tell us anything
about the viscosity.

As will become apparent below, a particularly interesting ratio to
consider is the ratio of shear viscosity and entropy density $s$.
The latter is proportional to $T^3$; thus
\begin{equation}
  \frac\eta s \sim \frac1{\lambda^2}\,.
\end{equation}
 One has $\eta/s \gg 1$ for $\lambda\ll1$.  
This is a common feature of weakly
coupled field theories.  Extrapolating to $\lambda\sim1$, one
finds $\eta/s\sim1$.  We shall see that
theories with gravity duals are strongly coupled, and $\eta/s$ is
of order one.  More surprisingly, this ratio is the same for all
theories with gravity duals.

To compute rather than estimate 
the viscosity, one can use
Kubo's formula.  It turns out that one has to sum an infinite number
of Feynman graphs to even find the viscosity to leading order.
Another way that leads to the same result is to first formulate a
kinetic Boltzmann equation for the quasi-particles as an intermediate
effective description, and then derive hydrodynamics by taking the
limit of very long lengths and time scales in the kinetic equation.
Interested readers should consult
Refs.~\cite{Jeon:1994if,Jeon:1995zm} for more details.

\section{AdS/CFT CORRESPONDENCE}
\label{sec:AdSCFT}

\subsection{Review Of AdS/CFT Correspondence At Zero Temperature}

This section briefly reviews the AdS/CFT correspondence at zero
temperature.  It contains only the minimal amount of materials
required to understand the rest of the review.  Further information can
be found in existing reviews and lecture
notes~\cite{Aharony:1999ti,Klebanov:2000me}.

The original example of AdS/CFT correspondence is between ${\cal N}=4$
supersymmetric Yang-Mills (SYM) theory and type IIB string theory on
AdS$_5\times$S$^5$ space.  Let us describe the two sides of the
correspondence in some more detail.

The ${\cal N}=4$ SYM theory is a gauge theory with a gauge field, four
Weyl fermions, and six real scalars, all in the adjoint representation
of the color group.  Its Lagrangian can be written down explicitly,
but is not very important for our purposes.  It has a vanishing beta
function and is a conformal field theory (CFT) (thus the CFT in
AdS/CFT).  In our further discussion, we frequently use the
generic terms ``field theory'' or CFT for the ${\cal N}=4$ SYM theory.

On the string theory side, we have type IIB string theory, which
contains a finite number of massless fields, including the graviton,
the dilaton $\Phi$, some other fields (forms) and their
fermionic superpartners, and an infinite number of massive string
excitations.  It has two parameters: the string length $l_s$ (related
to the slope parameter $\alpha'$ by $\alpha'=l_s^2$) and the string
coupling $g_s$.  In the long-wavelength limit, when all fields vary over
length scales much larger than $l_s$, the massive modes decouple and
one is left with type IIB supergravity in 10 dimensions, which can be
described by an action~\cite{Polchinski}
\begin{equation}\label{H-E}
  S_{\rm SUGRA} = \frac1{2\kappa_{10}^2}
  \int\!d^{10}x\,\sqrt{-g}\, e^{-2\Phi} \left({\cal R} 
  + 4\,\d^\mu\Phi\d_\mu\Phi +\cdots\right)\,,
\end{equation}
where $\kappa_{10}$ is the 10-dimensional gravitational constant,
\begin{equation}\label{kappa10}
  \kappa_{10} = \sqrt{8\pi G}= 8\pi^{7/2}g_s l_s^4\,,
\end{equation}
and $\cdots$ stay for the contributions from fields other than the
metric and the dilaton.  One of these fields is the five-form $F_5$,
which is constrained to be self-dual.  The type IIB string theory
lives is a 10-dimensional space-time with the following metric:
\begin{equation}\label{AdS-r}
  ds^2 = \frac{r^2}{R^2}(-dt^2+d\x^2) + \frac{R^2}{r^2}dr^2
         + R^2 d\Omega_5^2\,.
\end{equation}
The metric is a direct product of a five-dimensional sphere
($d\Omega_5^2$) and another five-dimensional space-time spanned by $t$,
$\x$, and $r$.  
An alternative form of the metric is obtained from Eq.~(\ref{AdS-r})
by a change of variable $z=R^2/r$,
\begin{equation}\label{AdS-z}
  ds^2 = \frac{R^2}{z^2}(-dt^2+d\x^2+dz^2) + R^2 d\Omega_5^2\,.
\end{equation}
Both coordinates $r$ and $z$ are known as the radial
coordinate.  The limiting value $r=\infty$ (or $z=0$) is the
boundary of the AdS space.

It is a simple exercise to check that the $(t,\x,r)$ part of the
metric is a space with constant negative curvature, or an anti
de-Sitter (AdS) space.  To support the metric~(\ref{AdS-r}) (i.e., to
satisfy the Einstein equation) there must be some background matter
field that gives a stress-energy tensor in the form of a negative
cosmological constant in AdS$_5$ and a positive one in S$^5$.  Such
a field is the self-dual five-form field $F_5$ mentioned above.

Field theory has two parameters: the number of colors $N$ and the
gauge coupling $g$.  When the number of colors is large, it is the 't
Hooft coupling $\lambda=g^2N$ that controls the perturbation theory.
On the string theory side, the parameters are $g_s$, $l_s$, and
radius $R$ of the AdS space.  String theory and field theory
each have two dimensionless parameters which map to each other through
the following relations:
\begin{eqnarray}
  g^2 &=& 4\pi g_s,\label{ggs} \\
  g^2N_c &=& \frac{R^4}{l_s^4}\,. \label{lambda-Rls}
\end{eqnarray}
Equation~(\ref{ggs}) tells us that, if one wants to keep
string theory weakly interacting, then the gauge coupling in field
theory must be small.  Equation~(\ref{lambda-Rls}) is
particularly interesting.  It says that the large 't Hooft coupling
limit in field theory corresponds to the limit when the curvature
radius of space-time is much larger than the string length $l_s$.  In
this limit, one can reliably decouple the massive string modes and
reduce string theory to supergravity.  In the limit $g_s\ll1$ and
$R\gg l_s$, one has classical supergravity instead of string theory.
The practical utility of the AdS/CFT correspondence comes, in large
part, from its ability to deal with the strong coupling limit in gauge
theory.

One can perform a Kaluza-Klein reduction~\cite{KK} by expanding all
fields in S$^5$ harmonics.  Keeping only the lowest harmonics, one
finds a five-dimensional theory with the massless dilaton, 
SO(6) gauge bosons, and
gravitons~\cite{Kim:1985ez}:
\begin{equation}\label{H-E-5D}
  S_{\rm 5D} = \frac{N^2}{8\pi^2R^3}\int\!d^5x\, 
    \left({\cal R}_{\rm 5D}-2\Lambda -\frac12
 \d^\mu\Phi\d_\mu\Phi 
   -\frac{R^2}8 F^a_{\mu\nu}F^{a\mu\nu} + \cdots
   \right)\,.
\end{equation}

In AdS/CFT,  an operator $O$ of field theory is put
in a correspondence with a field $\phi$ (``bulk'' field) in supergravity.
We elaborate on this correspondence below; here we
keep the operator and the field unspecified.  In the supergravity
approximation, the mathematical statement of the correspondence is
\begin{equation}\label{prescr}
  Z_{\rm 4D}[J] = e^{iS[\phi_{\rm cl}]}\,.
\end{equation}
On the left is the partition function of a field theory, where
the source $J$ coupled to the operator $O$ is included:
\begin{equation}
   Z_{\rm 4D}[J] = \int\!D\phi\,\exp\left(iS + i\!\int\!d^4x\, J O\right)\,.
\end{equation}
On the right, $S[\phi_{\rm cl}]$ is the classical action of
the  classical solution $\phi_{\rm cl}$ to the field equation with the
boundary condition:
\begin{equation}\label{bound-cond}
  \lim_{z\to0} \frac{\phi_{\rm cl}(z,x)}{z^\Delta} = J(x)\,.
\end{equation}
Here $\Delta$ is a constant that depends on the nature of the operator
$O$ (namely, on its spin and dimension).  In the  simplest case,
$\Delta=0$, and the boundary condition becomes $\phi_{\rm cl}(z{=}0)=J$.
Differentiating Eq.~(\ref{prescr}) with respect to $J$, one can find
the correlation functions of $O$.  For example, the two-point Green's
function of $O$ is obtained by differentiating $S_{\rm cl}[\phi]$
twice with respect to the boundary value of $\phi$,
\begin{equation}
  G(x-y) = -i \< T O(x) O(y)\> 
  = \left. - \frac{\delta^2 S[\phi_{\rm cl}]}{\delta J(x)\delta J(y)}
  \right|_{\phi(z=0)=J}\,.
\end{equation}
The AdS/CFT correspondence thus maps the problem of finding quantum
correlation functions in field theory to a classical problem in
gravity.  Moreover, to find two-point correlation functions in field
theory, one can be limited to the quadratic part of the classical
action on the gravity side.

The complete operator to field mapping can be found in
Refs.~\cite{Witten:1998qj,Aharony:1999ti}.  For our purpose, the
following is sufficient:
\begin{itemize}
\item  The dilaton $\Phi$ corresponds to $O=-{\cal L}=\frac14
F_{\mu\nu}^2+\cdots$, where ${\cal L}$ is the Lagrangian density.
\item The gauge field $A_\mu^a$ corresponds to the conserved R-charge
current $J^{a\mu}$ of field theory.
\item The metric tensor corresponds to the stress-energy tensor 
$T^{\mu\nu}$.  More precisely, the partition function of 
the four-dimensional field
theory in an external metric $g^0_{\mu\nu}$ is equal to 
\begin{equation}
  Z_{\rm 4D}[g^0_{\mu\nu}] = \exp(iS_{\rm cl}[g_{\mu\nu}])\,,
\end{equation}
where the five-dimensional metric $g_{\mu\nu}$ satisfies 
the Einstein's equations and has
the following asymptotics at $z=0$:
\begin{equation}
  ds^2 = g_{\mu\nu}dx^\mu dx^\nu = \frac{R^2}{z^2} 
     (dz^2 + g^0_{\mu\nu}dx^\mu dx^\nu)\,.
\end{equation}
\end{itemize}
From the point of view of hydrodynamics, the operator $\frac14F^2$ is
not very interesting because its correlator does not have a hydrodynamic
pole.  In contrast, we  find the correlators of the R-charge
current and the stress-energy tensor to contain hydrodynamic
information.  

We simplify the graviton part of the action further.  Our
two-point functions are functions of the momentum $p=(\omega,\k)$.
We can choose spatial coordinates so that $\k$ points along the 
$x^3$-axis.  This corresponds to perturbations that propagate along the
$x^3$ direction: $h_{\mu\nu}=h_{\mu\nu}(t,r,x^3)$.  These
perturbations can be classified according to 
 the  representations of the O(2) symmetry
of the $(x^1,x^2)$ plane.  Owing to that symmetry, only certain
components can mix; for example, $h_{12}$ does not mix with any other
components, whereas components $h_{01}$ and $h_{31}$ mix only with each
other.  We assume that only these three metric components are
nonzero and introduce shorthand notations
\begin{equation}
  \phi = h^1_2, \qquad a_0 = h^1_0,\qquad a_3 = h^1_3\,.
\end{equation}
The quadratic part of the graviton action acquires a very simple form
in terms of these fields:
\begin{equation}\label{S-phiAA}
  S_{\rm quad} = \frac{N^2}{8\pi^2R^3}\int\!d^4x\,dr\,\sqrt{-g}
  \left(-\frac12g^{\mu\nu}\d_\mu\phi\d_\nu\phi 
  -\frac1{4g_{\rm eff}^2} g^{\mu\alpha}g^{\nu\beta} 
  f_{\mu\nu}f_{\alpha\beta}\right)\,,
\end{equation}
where $f_{\mu\nu}=\d_\mu a_\nu-\d_\nu a_\mu$, and $g_{\rm
eff}^2=g_{xx}$.  In deriving Eq.~(\ref{S-phiAA}), our only assumption
about the metric is that it has a diagonal form,
\begin{equation}\label{metric-diag}
  ds^2 = g_{tt}dt^2 + g_{rr}dr^2 + g_{xx}d\x^2\,,
\end{equation}
so it can also be used below for the finite-temperature
metric.

As a simple example, let us compute the two-point correlation
function of $T^{xy}$, which corresponds to $\phi$ in gravity. 
The field equation for $\phi$ is
\begin{equation}
  \d_\mu(\sqrt{-g}\,g^{\mu\nu}\d_\nu\phi)=0\,. 
\end{equation}
The solution to this equation, with the boundary condition
$\phi(p,z=0)=\phi_0(p)$, can be written as
\begin{equation}
  \phi(p,z) = f_p(z)\phi_0(p)\,,
\end{equation}
where the mode function $f_p(z)$ satisfies the equation
\begin{equation}\label{fp-eq}
  \left(\frac{f_p'}{z^3}\right)' - \frac{p^2}{z^3}f_p =0
\end{equation}
with the boundary condition $f_p(0)=1$.  
The mode equation
(\ref{fp-eq}) 
can be solved exactly.  Assuming $p$ is spacelike,
$p^2>0$, the exact solution and its expansion around $z=0$ is
\begin{equation}
  f_p(z) = \frac12(pz)^2 K_2(pz) = 1 - \frac14(pz)^2 
  -\frac1{16} (pz)^4\ln(pz) + O((pz)^4)\,.
\end{equation}
The second solution to Eq.~(\ref{fp-eq}), $(pz)^2I_2(pz)$, is
ruled out because it blows up at $z\to\infty$.

We now substitute the solution into the quadratic action.  
Using the field
equation, one can perform integration by parts and write the action as
a boundary integral at $z=0$.  One finds
\begin{equation}
  S = \frac{N^2}{16\pi^2}\int\! d^4x\, 
      \frac1{z^3}\phi(x,z)\phi'(x,z)|_{z\to0}
  = \int\! \frac{d^4p}{(2\pi)^4}\, 
   \phi_0(-p){\cal F}(p,z) \phi_0(p)|_{z\to0}\,,
\end{equation}
where 
\begin{equation}
  {\cal F}(p,z) = \frac{N^2}{16\pi^2}\frac1{z^3} f_{-p}(z)\d_z f_p(z)\,.
\end{equation}
Differentiating the action twice with respect to the boundary value
$\phi_0$ one finds
\begin{equation}
  \< T_{xy} T_{xy}\>_p = -2 \lim_{z\to 0}{\cal F}(p,z)
  = \frac{N^2}{64\pi^2}p^4\ln(p^2)\,.
\end{equation}
Note that we have dropped the term $\sim p^4\ln z$, which,
although singular in the
limit $z\to0$, is a contact term [i.e., a term proportional to 
a derivative of $\delta(x)$ after Fourier transform].
Removing such terms by adding local counter terms 
to the supergravity action is known as the holographic 
renormalization \cite{Bianchi:2001kw}.
It is, in a sense, a holographic counterpart to the standard renormalization 
procedure in quantum field theory, here applied to composite operators.

For time-like $p$, $p^2<0$, there are two solutions to Eq.~(\ref{fp-eq})
which involve Hankel functions $H^{(1)}(z)$ and $H^{(2)}(z)$ instead of
$K_2(z)$. Neither function blows up at $z\to\infty$, and it is not clear
which should be picked.  Here we encounter, for the first time,
a subtlety of Minkowski-space AdS/CFT, which is discussed in
great length in subsequent sections.  At zero temperature this problem
can be overcome by an analytic continuation from space-like $p$.  However,
this will not work at nonzero temperatures.

\subsection{Black Three-Brane Metric}

At nonzero temperatures, the metric dual to ${\cal N}=4$ SYM theory is
the black three-brane metric,
\begin{equation}\label{finite-T-metric}
  ds^2 =  \frac{r^2}{R^2}(-fdt^2+d\x^2) + \frac{R^2}{r^2f}dr^2
         + R^2 d\Omega_5^2\,,
\end{equation}
with $f=1-r_0^4/r^4$.  The event horizon is located at $r=r_0$, where
$f=0$.  In contrast to the usual Schwarzschild black hole, the horizon
has three flat directions $\x$.  The metric~(\ref{finite-T-metric}) is
thus called a black three-brane metric.

We frequently use an alternative radial coordinate $u$,
defined as $u=r_0^2/r^2$.  In terms of $u$, the boundary is at $u=0$,
the horizon at $u=1$, and the metric is
\begin{equation}\label{finite-T-metric-u}
  ds^2 = \frac{(\pi TR)^2}{u^2}(-f(u)dt^2+d\x^2) + 
  \frac{R^2}{4u^2f(u)}du^2 + R^2 d\Omega_5^2\,.
\end{equation}

The Hawking temperature is determined completely by the behavior of
the metric near the horizon.  Let us concentrate on the $(t,r)$ part
of the metric,
\begin{equation}
  ds^2 = -\frac{4r_0}{R^2}(r-r_0)dt^2 + \frac{R^2}{4r_0(r-r_0)}dr^2\,.
\end{equation}
Changing the radial variable from $r$ to $\rho$,
\begin{equation}
  r = r_0 + \frac{\rho^2}{r_0}\,,
\end{equation}
and the metric components become nonsingular:
\begin{equation}
  ds^2 = \frac{R^2}{r_0^2}\left( d\rho^2 
  - \frac{4r_0^2}{R^2} \rho^2 dt^2\right)\,.
\end{equation}
Note also that after a Wick rotation to Euclidean time $\tau$,
the metric has the form of the flat metric in cylindrical coordinates,
$ds^2\sim d\rho^2+\rho^2d\varphi^2$, where $\varphi=2r_0R^{-2}\tau$.
To avoid a conical singularity at $\rho=0$, $\varphi$ must be
a periodic variable with periodicity $2\pi$.  This fact matches with the
periodicity of the Euclidean time in thermal field theory
$\tau\sim\tau+1/T$, from which one finds the Hawking temperature:
\begin{equation}\label{T_H}
  T_H = \frac{r_0}{\pi R^2}\,.
\end{equation}

One of the first finite-temperature predictions of AdS/CFT
correspondence is that of the thermodynamic potentials of the ${\cal
N}=4$ SYM theory in the strong coupling regime.  The entropy 
is given by the Bekenstein-Hawking formula $S=A/(4G)$, where
$A$ is the area of the horizon of the metric~(\ref{finite-T-metric}); 
the result can then be converted to
parameters of the gauge theory using Eqs.~(\ref{kappa10}), (\ref{ggs}),
and (\ref{lambda-Rls}).  One obtains
\begin{equation}\label{thermo}
  s =\frac SV = \frac{\pi^2}2N^2T^3\,,
\end{equation}
which is 3/4 of the entropy density in ${\cal N}=4$ SYM theory at zero 
't Hooft coupling.

We now try to generalize the AdS/CFT prescription to finite
temperature.  In the Euclidean formulation of finite-temperature field
theory, field theory lives in a space-time with the Euclidean time
direction $\tau$ compactified.  The metric is
regular at $r=r_0$: If one views the $(\tau, r)$ space as a
cigar-shaped surface, then the horizon $r=r_0$ is the tip of the cigar.
Thus, $r_0$ is the minimal radius where the space ends, and 
there is no point
in space with $r$ less than $r_0$.  The only boundary condition
at $r=r_0$ is that fields are regular at the tip of the cigar, and the
AdS/CFT correspondence is formulated as
\begin{equation}
  Z_{\rm 4D}[J] = Z_{\rm 5D}[\phi]|_{\phi(z=0)\to J}\,.
\end{equation}

\section{REAL-TIME AdS/CFT}
\label{sec:realtime}

In many cases we must find real-time correlation functions
not given directly by the Euclidean path-integral formulation of
thermal field theory.  One example is the set of 
kinetic coefficients expressed, through Kubo's formulas, 
via a certain limit of
real-time thermal Green's functions.  Another related example appears if
we want to directly find the position of the poles in the correlation
functions that would correspond to the hydrodynamic modes.

In principle, some real-time Green's functions can be obtained by
analytic continuation of the Euclidean ones.  For example, an analytic
continuation of a two-point Euclidean propagator gives a retarded or
advanced Green's function, depending on the way one performs the
continuation.  However, it is often very difficult to directly compute
a quantity of interest in that way.  In particular, it is very difficult
to get the information about the hydrodynamic (small $\omega$, small
$\k$) limit of real-time correlators from Euclidean propagators.  The
problem here is that we need to perform an analytic continuation from a
discrete set of points in Euclidean frequencies (the Matsubara
frequencies) $\omega=2\pi i n$, where $n$ is an integer, to the real values
of $\omega$.  In the hydrodynamic limit, we are interested in real and
small $\omega$, whereas the smallest Matsubara frequency is already
$2\pi T$.

Therefore, we need a real-time AdS/CFT prescription that would allow us 
to directly compute the real-time correlators.  However, if one tries to
naively generalize the AdS/CFT prescription, one immediately faces
 a problem.  Namely, now $r=r_0$ is not the end of space but
just the location of the horizon.  Without specifying a boundary
condition at $r=r_0$, there is an ambiguity in defining the solution to
the field equations, even as the boundary condition at $r=\infty$ is
set.

As an example, let us consider the equation of motion of a scalar
field in the black hole background, $\d_\mu (g^{\mu\nu}\d_\nu\phi)=0$.
The solution to this equation with the boundary condition
$\phi=\phi_0$ at $u=0$ is $\phi(p,u)=f_p\phi_0(p)$, where $f_p(u)$
satisfies the following equation in the
metric~(\ref{finite-T-metric-u}):
\begin{equation}\label{dilaton-T}
  f_p'' - \frac{1+u^2}{uf}f_p' + \frac{w^2}{uf^2}f_p
  - \frac{q^2}{uf}f_p =0\,.
\end{equation}
Here the 
prime denotes differentiation with respect to $u$, and we have
defined the dimensionless frequency and momentum:
\begin{equation}
  w = \frac\omega{2\pi T}\,, \qquad q = \frac k{2\pi T}\, .
\end{equation}
Near $u=0$ the equation has two solutions, $f_1\sim 1$ and $f_2\sim
u^2$.  In the Euclidean version of thermal AdS/CFT, there is only one
regular solution at the horizon $u=1$, which corresponds to a particular
linear combination of $f_1$ and $f_2$.  However, in Minkowski space there
are two solutions, and both are finite near the horizon.  One solution
termed $f_p$ behaves as $(1-u)^{-iw/2}$, and the other is its
complex conjugate $f_p^*\sim(1-u)^{iw/2}$.  These two solutions
oscillate rapidly as $u\to1$, but the amplitude of the 
oscillations is constant.  Thus, the
requirement of finiteness of $f_p$ allows for any linear combination
of $f_1$ and $f_2$ near the boundary, which means that there is
no unique solution to Eq.~(\ref{dilaton-T}).

\subsection{Prescription For Retarded Two-Point Functions}

Physically, the two solutions $f_p$ and $f_p^*$ 
have very different  behavior.  
 Restoring the $e^{-i\omega t}$
phase in the wave function, one can write
\begin{eqnarray}
  e^{-i\omega t} f_p \sim  e^{-i\omega(t+r_*)}\,,\label{incoming}\\
  e^{-i\omega t} f_p^* \sim e^{i\omega(t-r_*)}\,,\label{outgoing}
\end{eqnarray}
where the coordinate
\begin{equation}
  r_* = \frac{\ln(1-u)}{4\pi T}
\end{equation}
was introduced so that Eqs.~(\ref{incoming}) and (\ref{outgoing}) looked
like plane waves. In fact, Eq.~(\ref{incoming}) corresponds to a wave
that moves toward the horizon (incoming wave) and Eq.~(\ref{outgoing})
to a wave that moves away from the horizon (outgoing wave).

The simplest idea, which is motivated by the fact that nothing should
come out of a horizon,
is to impose the incoming-wave boundary condition at
$r=r_0$ and then proceed as instructed by the AdS/CFT correspondence.
However, now we encounter another problem.  If we write down the
classical action for the bulk field, after integrating by parts we get
 contributions from both the boundary and the horizon:
\begin{equation}
  S = \int\!\frac{d^4p}{(2\pi)^4}\, 
  \phi_0(-p){\cal F}(p,z)\phi_0(p)\Bigl|^{z=z_H}_{z=0}\,.
\end{equation}
If one tried to differentiate the action with respect to the boundary
value $\phi_0$, one would find
\begin{equation}\label{Gp}
  G(p) = {\cal F}(p,z)|^{z_H}_0 + {\cal F}(-p,z)|^{z_H}_0\,.
\end{equation}
From the equation satisfied by $f_p$ and from $f_p^*=f_{-p}$, it is
easy to show that the imaginary part of ${\cal F}(p,z)$ does not
depend on $z$; hence the quantity $G(p)$ in Eq.~(\ref{Gp}) is real.
This is clearly not what we want, as the retarded Green's functions
are, in general, complex.  Simply throwing away the contribution from
the horizon does not help because ${\cal F}(-p,z)={\cal F}^*(p,z)$ owing
to the reality of the equation satisfied by $f_p$.

A partial solution to this problem was suggested in
Ref.~\cite{Son:2002sd}. It was postulated that the retarded Green's
function is related to the function ${\cal F}$ by the same formula
that was found at zero temperature:
\begin{equation}\label{GR-prescr}
  G^R(p) = -2 \lim_{z\to0} {\cal F}(p,z)\,.
\end{equation}
In particular, we throw away all contributions from the horizon.  This
prescription was established more rigorously in
Ref.~\cite{Herzog:2002pc} (following an earlier suggestion in
Ref.~\cite{Maldacena:2001kr}) as a particular case of a general
real-time AdS/CFT formulation, which establishes the connection
between the close-time-path formulation of real-time quantum field
theory with the dynamics of fields in the whole Penrose diagram of the
AdS black brane.  Here we accept Eq.~(\ref{GR-prescr}) as a postulate
and proceed to extract physical results from it.

It is also easy to generalize this prescription to the case when we have
 more than one field.  In that case, the quantity ${\cal F}$
becomes a matrix ${\cal F}_{ab}$, whose elements are proportional to
the retarded Green's function $G_{ab}$.

\subsection{Calculating Hydrodynamic Quantities}

As an illustration of the real-time AdS/CFT correspondence, we
compute the correlator of $T_{xy}$.  First we write down
the equation of motion for $\phi=h^x_y$:
\begin{equation}
  \phi_p'' - \frac{1+u^2}{uf} \phi_p' + \frac{w^2-q^2f}{uf^2}\phi_p = 0\,.
\end{equation}
In contrast to the zero-temperature equation, now $\omega$ and $k$
enter the equation separately rather than through the combination
$\omega^2-k^2$.  Thus the Green's function will have no Lorentz
invariance.  The equation cannot be solved exactly for all $\omega$ and
$k$.  However, when $\omega$ and $k$ are both much smaller than $T$, or
$w,q\ll1$,
one can develop series expansion in powers of $w$ and $q$.
There are two solutions that are complex conjugates of each
other.  The solution that is an incoming wave at $u=1$ and 
normalized to 1 at $u=0$ is
\begin{equation}
  f_p(z) = (1-u^2)^{-i w/2} + O(w^2,q^2)\,.
\end{equation}
The kinetic term in the action for $\phi$ is
\begin{equation}
  S = -\frac{\pi^2N^2T^4}8\int\!du\,\frac fu \phi'^2\,.
\end{equation}
Applying the general formula~(\ref{GR-prescr}), 
one finds the retarded Green's
function of $T_{xy}$,
\begin{equation}
  G^R_{xy,xy}(\omega,k) = -\frac{\pi^2N^2T^4}{4} i w \,,
\end{equation}
and, using Kubo's formula for $\eta$, the viscosity,
\begin{equation}\label{etaAdS}
  \eta = \frac\pi 8 N^2 T^3\,.
\end{equation}

It is instructive to compute other correlators that have poles
corresponding to  hydrodynamic modes.  As a warm-up, let us compute
the two-point correlators of the R-charge currents, which should have a
pole at $\omega=-iDk^2$, where $D$ is the diffusion constant.  We first
write down Maxwell's equations for the bulk gauge field.  Let
the spatial momentum be aligned along the $x^3$-axis: 
$p=(\omega,0,0,k)$.  Then
the equations for $A_0$ and $A_3$ are coupled:
\begin{eqnarray}
  w A_0' + q f A_3' &=& 0\,,\\
  A_0'' - \frac1{uf} (q^2 A_0 + wq A_3) &=& 0\,,\label{eq-A0}\\
  A_3'' + \frac{f'}f A_3' + \frac1{uf^2}(w^2A_3+wqA_0) &=& 0\,.
\label{eq-A3}
\end{eqnarray}
One can eliminate
$A_3$ and write down a third-order equation for $A_0$,
\begin{equation}\label{Atprime}
  A_0''' + \frac{(uf)'}{uf} A_0'' + \frac{w^2-q^2f}{uf^2}A_0' = 0\,.
\end{equation}
Near $u{=}1$ we find two independent solutions, $A_0'\sim(1-u)^{\pm
iw/2}$, and the incoming-wave boundary condition singles out
$(1-u)^{-iw/2}$.  One can substitute $A_0'=(1-u)^{-iw/2}F(u)$ into
Eq.~(\ref{Atprime}). The resulting
 equation can be solved perturbatively in $w$
and $q^2$. We find
\begin{equation}
  A_0' = C (1-u)^{-iw/2} \left( 1+ \frac{iw}2\ln\frac{2u^2}{1+u}
         + q^2\ln\frac{1+u}{2u} \right)\,.
\end{equation}
 Using Eq.~(\ref{eq-A0}) one can express $C$ through
 the boundary values
 of $A_0$ and $A_3$ at $u=0$:
\begin{equation}
  C = \left. \frac{q^2 A_0 + wq A_3}{iw-q^2}\right|_{u=0}\,.
\end{equation}
Differentiating the action with respect to the boundary values, we
find, in particular,
\begin{equation}\label{JJAdS}
  \<J_0 J_0\>_p = \frac{N^2T}{16\pi} \frac{k^2}{i\omega-Dk^2}\,,
\end{equation}
where
\begin{equation}
  D = \frac1{2\pi T}\,.
\end{equation}
The correlator given by Eq.~(\ref{JJAdS}) 
has the expected hydrodynamic diffusive pole,
and $D$ is the R-charge diffusion constant. 

Similarly, one can observe the appearance of the shear mode in the
correlators of the metric tensor.  We note that the shear flow along
the $x^1$ direction with velocity gradient along the $x^3$ direction
involves $T_{01}$ and $T_{31}$, hence the interesting metric
components are $a_0=h_0^1$ and $a_3=h_3^1$.  Two of the field
equations are
\begin{eqnarray}
  && a_0' - \frac{qf}w a_3' = 0\,,\\
  && a_3'' - \frac{1+u^2}{uf}a_3' + \frac1{uf^2}(w^2a_3+wqa_0) = 0\,.
\end{eqnarray}
They can be combined into a single equation:
\begin{equation}
  a_0''' - \frac{2u}f a_0'' + \frac{2uf-q^2f+w^2}{uf^2}a_0' = 0\,.
\end{equation}
Again, the solution can be found perturbatively in $w$ and $q$:
\begin{equation}
  a_0' = C (1-u)^{-iw/2} \left[ u- iw \left( 1-u-\frac u2
  \ln \frac{1+u}2\right) + \frac{q^2}2 (1-u)\right]\,.
\end{equation}
Applying the prescription, one finds the retarded Green's functions.
For example,
\begin{equation}
  G_{tx,tx}(\omega,k) = \frac{\xi k^2}{i\omega-{\cal D}k^2}\,,
\end{equation}
where
\begin{equation}
  \xi = \frac\pi8 N^2T^3, \qquad {\cal D}=\frac1{4\pi T}\,.
\end{equation}
Thus, we 
found that the correlator contains a diffusive pole $\omega=-i{\cal
D}k^2$, just as anticipated from hydrodynamics.  Furthermore, the
magnitude of the momentum diffusion constant ${\cal D}$ also matched
 our expectation.  Indeed, if one recalls the value of $\eta$ from 
Eq.~(\ref{etaAdS}) and the entropy density from Eq.~(\ref{thermo}), one can
check that
\begin{equation}
  {\cal D} = \frac\eta{\epsilon+P}\,.
\end{equation}

\section{THE MEMBRANE PARADIGM}
\label{sec:membrane}

Let us now look at the problem from a different perspective.
The existence of 
hydrodynamic modes in thermal field theory is reflected by 
the existence of 
the poles of the retarded 
correlators computed from gravity.  Are there direct
gravity counterparts of the hydrodynamic normal modes?

If the answer to this question is yes, then there must exist linear
gravitational perturbations of the metric that have the dispersion
relation identical to that of the shear hydrodynamic mode,
$\omega\sim-iq^2$, and of the sound mode, $\omega=c_s q-i\gamma q^2$.
It turns out that one can explicitly construct the gravitational
counterpart of the shear mode.  (It should be possible to find
a similar construction for the sound mode, but it has not been done in the
literature; for a recent work on the subject,
 see \cite{Saremi:2007dn}.)  Our discussion is physical but somewhat sketchy;
for more details see Ref.~\cite{Kovtun:2003wp}.

First, let us construct a gravity perturbation that
corresponds to a diffusion of a conserved charge (e.g., the R-charge in
${\cal N}=4$ SYM theory).  To keep the discussion general, we use 
the form of the metric~(\ref{metric-diag}), with the metric components
 unspecified. Our only assumptions are that the metric is diagonal and 
   has a horizon
at $r=r_0$, near which
\begin{equation}
  g_{00} = -\gamma_0 (r-r_0), \qquad g_{rr} = \frac{\gamma_r}{r-r_0}\,.
\end{equation}
The Hawking temperature can be computed by the method used to arrive
at Eq.~(\ref{T_H}), and one finds
$T=(4\pi)^{-1}(\gamma_0/\gamma_r)^{1/2}$.

We also assume that the action of the gauge field dual to the
conserved current is
\begin{equation}
  S_{\rm gauge} = \int\!dx\, \sqrt{-g}\left(-\frac1{4g_{\rm eff}^2}
    F^{\mu\nu} F_{\mu\nu}\right)\,,
\end{equation}
where $g_{\rm eff}$ is an effective gauge coupling that can be a
function of the radial coordinate $r$.  For simplicity we set $g_{\rm
eff}$ to a constant in our derivation of the formula for $D$; it can be
restored by replacing $\sqrt{-g}\to\sqrt{-g}/g_{\rm eff}^2$ in the
final answer.

The field equations are
\begin{equation}\label{Maxwell-gen}
  \d_\mu\left( \frac1{g_{\rm eff}^2}\sqrt{-g}\, F^{\mu\nu}\right) =0\,.
\end{equation}
We search for a solution to this equation that vanishes at
the boundary and satisfies the incoming-wave boundary condition at the horizon.

The first indication that one can have a hydrodynamic behavior on the
gravity side is that Eq.~(\ref{Maxwell-gen}) implies a conservation
law on a four-dimensional surface.  We define the stretched horizon 
as a surface
with constant $r$ just outside the horizon,
\begin{equation}
  r = r_h = r_0 + \varepsilon, \qquad \varepsilon\ll r_0\,,
\end{equation}
and the normal vector $n_\mu$ directed along the $r$ direction (i.e.,
perpendicularly to the stretched horizon).  Then with any solution to
Eq.~(\ref{Maxwell-gen}), one can associate a current on the stretched
horizon:
\begin{equation}
  j^\mu = n_\nu F^{\mu\nu}\Big|_{r_h}\,.
\end{equation}
The antisymmetry of $F^{\mu\nu}$ implies that $j^\mu$ has no radial
component, $j^r=0$.  The field equation (\ref{Maxwell-gen}) and the
constancy of $n_\nu$ on the stretched horizon imply that this
current is conserved: $\d_\mu j^\mu = 0$.  To establish the diffusive
nature of the solution, we must show the validity of the
constitutive equation $j^i=-D\d_i j^0$.

Such constitutive equation breaks time reversal and obviously must
come from the absorptive boundary condition on the horizon.  The
situation is analogous to the propagation of plane waves to a
non-reflecting surface in classical electrodynamics.  In this case, we
have the relation ${\bf B}=-{\bf n}\times{\bf E}$ between electric
and magnetic fields.  In our case, the corresponding relation is
\begin{equation}
  F_{ir} = -\sqrt{\frac{\gamma_r}{\gamma_0}}\, \frac{F_{0i}}{r-r_0}\,,
\end{equation}
valid when $r$ is close to $r_0$.  This relates $j_i\sim F_{ir}$ to
the parallel to the horizon component of the electric field
$F_{0i}$, which is one of the main points of the ``membrane paradigm''
approach to black hole physics~\cite{Damour:1978cg, paradigm}.  
We have yet to
relate $j_i$  to $j_0\sim F_{0r}$, which is the component of the
electric field normal to the horizon.  To make the connection to
$F_{0r}$, we use the radial gauge $A_r=0$, in which
\begin{equation}
  F_{0i} \approx - \d_i A_0\,.
\end{equation}
Moreover, when $k$ is small the fields change very slowly along the
horizon.  Therefore, at each point on the horizon the radial dependence
of the scalar potential $A_0$ is determined by the Poisson equation,
\begin{equation}
  \d_r (\sqrt{-g}\, g^{rr}g^{00}\d_r A_0 ) = 0\,,
\end{equation}
whose solution, which satisfies $A_0(r=\infty)=0$, is
\begin{equation}\label{A0r}
  A_0(r) = C_0 \int\limits_r^\infty\! dr'\, 
  \frac{g_{00}(r')g_{rr}(r')}{\sqrt{-g(r')}}\,.
\end{equation}
This means that the ratio of the scalar potential $A_0$ and electric
field $F_{0r}$ approaches a constant near the horizon:
\begin{equation}\label{A0F0r}
  \left. \frac{A_0}{F_{0r}}\right|_{r=r_0} =
  \frac{\sqrt{-g}}{g_{00}g_{rr}}(r_0)\int\limits_{r_0}^\infty\!
  dr\, \frac{g_{00}g_{rr}}{\sqrt{-g}}(r)\,.
\end{equation}
Combining the formulas $j^i\sim F_{0i}\sim\d_i A_0$, and
$A_0\sim F_{0r}\sim j^0$, we find Fick's law $j^i=-D\d_i j^0$, with
the diffusion constant
\begin{equation}\label{D-general}
  D = \frac{\sqrt{-g}}{g_{xx}g^2_{\rm eff}\sqrt{-g_{00}g_{rr}}}(r_0)
  \int\limits_{r_0}^\infty\!dr\,
\frac{-g_{00}g_{rr}g^2_{\rm eff}}{\sqrt{-g}}(r)\,.
\end{equation}

Thus, we found that for a slowly varying solution to Maxwell's
equations, the corresponding charge on the stretched horizon evolves
according to the diffusion equation.  Therefore, the gravity solution
must be an overdamped one, with $\omega=-iDk^2$.  This is an example of
a quasi-normal mode.  We also found the diffusion constant $D$ directly
in terms of the metric and the gauge coupling $g_{\rm eff}$.

The reader may notice that our quasinormal modes satisfy a vanishing
 Dirichlet condition at the boundary $r{=}\infty$.  This is different
from the boundary condition one uses to find the retarded
propagators in AdS/CFT, so the relation of the quasinormal modes to
AdS/CFT correspondence may be not clear.  It can be shown, however,
that the quasi-normal frequencies coincide with the poles of the retarded
correlators~\cite{Kovtun:2005ev,andrei}.

We can now apply our general formulas to the case of ${\cal N}=4$ SYM
theory.  The metric components are given by Eq.~(\ref{finite-T-metric}).
For the R-charge current $g_{\rm eff}=\textrm{const}$,
Eq.~(\ref{D-general}) gives $D=1/(2\pi T)$, in agreement with our
AdS/CFT computation.  For the shear mode of the stress-energy tensor
we have effectively $g^2_{\rm eff}=g_{xx}$, so ${\cal D}=1/(4\pi T)$,
which also coincides with our previous result.  In both cases, the
computation is much simpler than the AdS/CFT calculation.

\section{THE VISCOSITY/ENTROPY RATIO}
\label{sec:ratio}

\subsection{Universality}

In all thermal field theories in the regime
 described by gravity duals 
the ratio of shear viscosity $\eta$ 
to (volume) density of entropy $s$ is a universal 
constant equal to $1/(4\pi)$ [$\hbar/(4\pi k_B)$,
 if one restores $\hbar$, $c$ and the Boltzmann constant $k_B$].

One proof of the universality is based on the relationship between
graviton's absorption cross section and the imaginary part of the
retarded Green's function for $T_{xy}$ \cite{Kovtun:2004de}.  
Another way to prove the universality \cite{Buchel:2004qq} 
is via the direct
AdS/CFT calculation of the correlation function in Kubo's formula
(\ref{kubo}).

We, however, follow a
different method.  It is based on the formula for the viscosity
derived from the membrane paradigm.  A similar proof was
given by Buchel \& Liu~\cite{Buchel:2003tz}.  
%

The observation is that the shear gravitational perturbation with $k=0$
can be found exactly by performing a Lorentz boost of the black-brane
metric~(\ref{finite-T-metric}). 
Consider the coordinate transformations $r,t,x_i \rightarrow
r',t',x_i'$ of the form
\begin{eqnarray}
r &=& r'\,, \nonumber \\
t &=& \frac{t' + v y'}{\sqrt{1-v^2}}\approx t' + v y'\,, \nonumber \\
y &=&  \frac{y' + v t'}{\sqrt{1-v^2}}\approx y' + v t'\,, \nonumber \\
x_i &=& x_i'\,,
\label{transf}
\end{eqnarray}
where $v<1$ is a constant parameter and the expansion 
on the right corresponds to $v\ll 1$. 
In the new coordinates, the metric becomes
\begin{equation}
ds^2 =  g_{00}dt'\,^2 + g_{rr}dr'\,^2 + g_{xx}(r) \sum_{i=1}^p (dx'\,^i)^2
+ 2 v (  g_{00} +  g_{xx} ) dt' dy'\,. 
\label{newg}
\end{equation}
This is simply a shear fluctuation at $k=0$.  In our language, the
corresponding gauge potential is
\begin{equation}\label{a0g}
  a_0 = v g^{xx}(g_{00}+g_{xx})\,.
\end{equation}
This field satisfies the vanishing boundary condition
$a_0(r=\infty)=0$ owing to the restoration of Poincar\'e invariance at the
boundary: $g_{00}/g_{xx}\to-1$ when $r\to\infty$.  This clearly has a
much simpler form than Eq.~(\ref{A0r}) for the solution to the generic
Poisson equation.  The simple form of solution~(\ref{a0g}) is
valid only for the specific case of the shear gravitational mode with $g_{\rm
eff}^2=g_{xx}$.  We have also implicitly used the fact that the metric
satisfies the Einstein equations, with the stress-energy tensor on the
right being invariant under a Lorentz boost.

Equation~(\ref{A0F0r}) now becomes
\begin{equation}
\frac{a_0}{f_{0r}}\Biggl|_{r\rightarrow r_0} = - \frac{1+g^{xx}g_{00}}
{\partial_r (g^{xx}g_{00})}\Biggl|_{r\rightarrow r_0} = \frac{g_{xx}(r_0)}{
\gamma_0}\,.
\label{missing}
\end{equation}
The shear mode diffusion constant is
\begin{equation}
 {\cal D} =  \frac{a_0}{f_{0r}}\Biggl|_{r\rightarrow r_0} 
  \frac{\sqrt{\gamma_0 \gamma_r}}{g_{xx}(r_0)} =
 \sqrt{\frac{\gamma_r}{\gamma_0}} = \frac{1}{4\pi T}\,.
\end{equation}
Because ${\cal D}=\eta/(\epsilon+P)$, and $\epsilon+P=Ts$ in the absence
of chemical potentials, we find that
\begin{equation}
  \frac\eta s = \frac1{4\pi}\,.
\end{equation}
In fact, the constancy of this ratio has been checked directly 
for theories dual to $Dp$-brane \cite{Kovtun:2003wp},
 $M$-brane \cite{Herzog:2002fn}, Klebanov-Tseytlin
and Maldacena-Nunez backgrounds \cite{Buchel:2003tz},
 ${\cal N}=2^*$ SYM theory 
\cite{Buchel:2004hw} and others.
 Curiously, the viscosity to entropy ratio
is also equal to $1/4\pi$ in the pre-AdS/CFT ``membrane paradigm''
hydrodynamics \cite{damour}: there, for a four-dimensional
Schwarzschild black hole one has $\eta_{\mbox{{\tiny m.p.}}}
 = 1/16 \pi G_N$,
 while the Bekenstein-Hawking entropy is $s=1/4G_N$.

As remarked in Sec.~\ref{sec:hydro}, the ratio $\eta/s$ is much larger
than the one for weakly coupled theories. The fact that 
 we found the ratio to be
parametrically of order one implies that all theories with gravity
duals are strongly coupled.

In ${\cal N}=4$ SYM theory, the ratio $\eta/s$ has been computed to
the next order in the  inverse 't Hooft 
coupling expansion~\cite{Buchel:2004di}
\begin{equation}
  \frac\eta s = \frac1{4\pi} \left( 1+ \frac{135\zeta(3)}{8(g^2N)^{3/2}}
  \right)\,.
\end{equation}
The sign of the correction can be guessed from the fact that in the
limit of zero 't Hooft coupling $g^2N\to0$, the ratio diverges,
$\eta/s\to\infty$.

\subsection{The Viscosity Bound Conjecture}
From our discussion above, one can argue that 
\begin{equation}
  \frac\eta s \geq \frac{\hbar}{4\pi}
\end{equation}
in all systems that can be obtained from a sensible relativistic
quantum field theory by turning on temperatures and chemical
potentials.

The bound, if correct, implies that a liquid with a given volume density of
entropy cannot be arbitrarily close to being a perfect fluid (which
has zero viscosity).  As such, it implies a lower bound on the
viscosity of the QGP one may be creating at RHIC.
Interestingly, some model calculations suggest that the viscosity at RHIC 
may be not too far away from the lower 
bound~\cite{Teaney:2003kp,Shuryak:2003xe}.  

One place where one may think that the bound should break down is
superfluids.  The ability of a superfluid to flow without dissipation
in a channel is sometimes described as ``zero viscosity''. 
However, within the Landau's two-fluid model, any superfluid has a measurable
shear viscosity (together with three bulk viscosities).  For
superfluid helium, the shear viscosity has been measured in a
torsion-pendulum experiment by Andronikashvili~\cite{Andronikashvili}.  
If one substitutes the
experimental values, the ratio $\eta/s$ for helium remains larger
than $\hbar/4\pi k_B \approx 6.08\times 10^{-13}$ K$\,$s 
for all ranges of temperatures and pressures, by a
factor of at least 8.8.

As discussed in Sec.~\ref{visc:weak}, 
the ratio $\eta/s$ is proportional to the
ratio of the mean free path and the de Broglie wavelength of
particles,
\begin{equation}
  \frac\eta s \sim \frac{\ell_{\rm mfp}}{\lambda}\,.
\end{equation}
For the quasi-particle picture to be valid, the mean free path must be
much larger than the de Broglie wavelength.  Therefore, if the
coupling is weak and the system can be described as a collection of
quasi-particles, the ratio $\eta/s$ is larger than 1.

We have found is that, 
within the ${\cal N}=4$ SYM theory and, more generally,
 theories with gravity duals, even
in the limit of infinite coupling the ratio $\eta/s$ cannot be made
smaller than $1/(4\pi)$.

\section{CONCLUSION}

In this review, we covered only a small part of the applications
of AdS/CFT correspondence to finite-temperature quantum field theory.
Here we briefly mention further developments and refer the reader to
the original literature for more details.

In addition to ${\cal N}=4$ SYM theory, there exists a large
number of other theories whose hydrodynamic behavior has been
studied using the AdS/CFT correspondence, including the worldvolume
theories on M2- and M5-branes~\cite{Herzog:2002fn}, theories on D$p$
branes~\cite{Kovtun:2003wp}, and little string
theory~\cite{Parnachev:2005hh}.  In all examples the ratio $\eta/s$ is
equal to $1/(4\pi)$, which is not surprising because the general proofs of
Sec.~\ref{sec:ratio} apply in these cases.

We have concentrated on the shear hydrodynamic mode,
which has a diffusive pole ($\omega\sim -ik^2$).  One can also compute
correlators which have a sound-wave pole from the AdS/CFT
prescription \cite{Policastro:2002tn}. 
One such correlator is between the energy density
$T^{00}$ at two different points in space-time.  The result confirms
the existence of such a pole, with both the real part and imaginary part
having exactly the values predicted by hydrodynamics (recall
that in conformal field theories the bulk viscosity is zero and the
sound attenuation rate is determined completely by the shear
viscosity).

Some of the theories listed above are conformal field theories,
but many are not (e.g., the D$p$-brane worldvolume theories with
$p\neq3$).  The fact that $\eta/s=1/(4\pi)$ also in those theories 
implies that the constancy of
this ratio is not a consequence of conformal symmetry.  Theories with
less than maximal number of supersymmetries have been found to have the
universal value of $\eta/s$, for example, the ${\cal N}=2^*$
theory~\cite{Buchel:2003ah},  theories described by
Klebanov-Tseytlin, and Maldacena-Nunez backgrounds~\cite{Buchel:2003tz}.
A common feature of these theories is that they all
have a gravitational dual description.  The bulk viscosity has been
computed for some of these
theories~\cite{Benincasa:2005iv,Parnachev:2005hh}.

Besides viscosity, one can also compute diffusion constants of conserved
charges by using the AdS/CFT correspondence.  Above we presented the
computation of the R-charge diffusion constant in ${\cal N}=4$ SYM
theory; for similar calculations in some other theories see
Ref.~\cite{Herzog:2002fn,Kovtun:2003wp}.  

Recently, the AdS/CFT correspondence was used to compute the
 energy loss rate of
 a quark in the fundamental representation
moving in a finite-temperature
plasma~\cite{Herzog:2006gh,Liu:2006ug,Casalderrey-Solana:2006rq,Gubser:2006bz}.
This quantity is of importance to the phenomenon of ``jet quenching''
in heavy-ion collisions.

So far, the only quantity that shows a universal behavior at the
quantitative level, across all theories with gravitational duals, is the
ratio of the shear viscosity and entropy density.  Recently, it was
found that this ratio remains constant even at nonzero chemical
potentials \cite{Son:2006em,Mas:2006dy,Maeda:2006by,Saremi:2006ep,
Benincasa:2006fu}.

What have we learned from the application of AdS/CFT correspondence to
thermal field theory?  Although, at least at this moment, we cannot use
the AdS/CFT approach to study QCD directly, 
we have found quite interesting facts
about strongly coupled field theories.  We have also learned new facts
about quasi-normal modes of  black branes.  However, we have also
found a set of puzzles: Why is the ratio of the viscosity and entropy
density constant in a wide class of theories?  Is there a lower
bound on this ratio for all quantum field theories?  Can this be
understood without any reference to gravity duals?  With these open
questions, we conclude this review.

\bigskip

\noindent
{\sc Acknowledgments}

\noindent
The work of DTS is supported in part by U.S.\ Department of Energy
under Grant No.\ DE-FG02-00ER41132.  Research at Perimeter Institute
is supported in part by the Government of Canada through NSERC and by
the Province of Ontario through MEDT.

\end{document}